\documentclass[aps,prl,preprint,groupedaddress,showpacs]{revtex4}

\usepackage{graphicx}
\usepackage{graphics}
\usepackage{psfig}	
\usepackage{bm}
\begin{document}


\title{Dark Matter and Dark Energy from the solution of the
strong--CP problem}


\author{Roberto Mainini \& Silvio A. Bonometto}
\affiliation{Dipartimento di Fisica G.~Occhialini, Universit\`a di
Milano--Bicocca, Piazza della Scienza 3, 20126 Milano, Italy
\& I.N.F.N., Sezione di Milano}


\date{\today}

\begin{abstract}
The Peccei--Quinn (PQ) solution of the strong--$CP$ problem requires 
the existence of axions, which are a viable candidate for Dark Matter. 
Here we show that, if the Nambu--Goldstone potential of the PQ model
is replaced by a potential $V(|\Phi|)$ admitting a tracker solution, 
the scalar field $|\Phi|$ can account for Dark Energy, while the
phase of $\Phi$ yields axion Dark Matter. Such Dark Matter and Dark 
Energy turn out to be weakly coupled. If $V$ is a SUGRA
potential, the model essentially depends on a single parameter, the 
energy scale $\Lambda$. Once we set $\Lambda \simeq 10^{10}$GeV,
at the quark--hadron transition, $|\Phi|$ naturally passes through 
values suitable to solve the strong--$CP$ problem, later growing
to values providing fair amounts of Dark Matter and Dark Energy. 
In this model, the linear growth factor, from recombination to now, 
is quite close to $\Lambda$CDM. The selected $\Lambda$ value can be 
an indication of the scale where the soft breaking of SUSY occurred.

\end{abstract}


\maketitle

\section{Introduction}
The solutions of the strong $CP$ problem proposed by
Peccei \& Quinn in 1977  (\cite{1}, PQ hereafter) leads to one of the 
accepted models of Dark Matter (DM). PQ consider the 
lagrangian term
\begin{equation}
{\cal L}_\theta = {\alpha_s \over 2\pi} \theta \, G \cdot {\tilde G}
\label{eq:n1}
\end{equation}
($\alpha_s$: strong coupling constant, $G$ and $ {\tilde G}$: 
gluon field tensor and its dual), yielding $CP$~violations in strong 
interactions, and show that its effects are suppressed
by making $\theta$~a dynamical variable,~approaching zero 
in our cosmic era, its
residual oscillations appearing as DM \cite{2,3}.

The $\theta$ dynamics is set by assuming that a complex field
$\Phi =  \phi e^\theta/\sqrt{2}$ exists, whose evolution 
is ruled by a Nambu--Goldstone potential
\begin{equation}
V(|\Phi|) = \lambda [|\Phi|^2 - F_{PQ}^2]^2 ~,
\label{eq:n2}
\end{equation}
which is clearly $U(1)$ invariant. At $T<F_{PQ}$ (the PQ energy 
scale, which shall be $\sim 10^{12}$GeV), $\phi$ falls into the potential 
minimum, so that the $U(1)$ symmetry breaks, as $\theta$ acquires different 
values in different horizons. When the chiral symmetry is also broken,
close to the quark--hadron transition, a further term
must be added to the effective lagrangian, arising because of
instanton effects. This term reads
\begin{equation}
V_1 = \big[\sum_q \langle 0(T)| {\bar q} q |0(T) \rangle m_q \big] 
~(1 - \cos \theta) ~.
\label{eq:n3}
\end{equation}
At $T \simeq 0$, the square bracket approaches $m_\pi^2 f_\pi^2$ ($m_\pi$, 
$f_\pi$: $\pi$--meson mass, decay constant).
\vglue 0.05truecm

The choice of a NG potential is the simplest possible.
Here we explore the possibility of replacing 
it by a potential with a tracker solution \cite{4,5}.
Instead of taking a value $\simeq F_{PQ}$ soon, $\phi$ evolves over
cosmological times. As in the PQ case, the potential shall
involve a complex field $\Phi$ and be $U(1)$ invariant. 
While $\phi$ rapidly settles on the tracker solution (apart of 
residual fluctuations) in almost any horizon, the 
symmetry is broken soon by the values taken by $\theta$,
which suffers no dynamical constraints and is therefore
random, in different horizons. Later on,
when a mass term arises because of the chiral symmetry break,
dynamics becomes relevant also for the $\theta$ degree
of freedom, as in the PQ case. Here this happens while $\phi$ 
still evolves over cosmological times.
Finally, in the present epoch, $\phi$ accounts
for Dark Energy (DE). Hence, besides of yielding DM through
its phase $\theta$, the $\Phi$ field, introduced to solve the 
strong $CP$ problem, accounts for DE through its modulus $\phi$.

Within this model, DM and DE will be weakly coupled. If we take
a generalization of the SUGRA potential \cite{5} 
as tracking potential, with an energy scale $\Lambda \sim 10^{10}$GeV,
we find reasonable values for today's DM and DE densities,
while $\theta$ is driven to values even smaller than in the PQ case, so
that $CP$ is apparently conserved in strong interactions.
In turn, $\Lambda$ may be an indication of the scale where the
soft breaking of super--symmetries occurred.

\section{Lagrangian theory}
The lagrangian ${\cal L} =  \sqrt{-g} \{ g_{\mu\nu} 
\partial_\mu \Phi \partial_\nu \Phi   - V(|\Phi|) \} $
can be rewritten in terms of $\phi$ and $\theta$, adding
also the term breaking the $U(1)$ symmetry, as follows:
\begin{eqnarray}
{\cal L} = \sqrt{-g} \{ {1 \over 2} 
g_{\mu\nu} [\partial_\mu \phi \partial_\nu \phi 
+ \phi^2  \partial_\mu \theta \partial_\nu \theta] 
- V(\phi) -m^2(T,\phi) \phi^2 (1 - \cos \theta) 
\}  ~.
\label{eq:m1}
\end{eqnarray}	
Here $g_{\mu\nu}$ is the metric tensor. We shall assume that
$ds^2 = g_{\mu\nu} dx^\mu dx^\nu = 
a^2 (d\tau^2 - \eta_{ij}dx_idx_j)$, so that $a$ is the
scale factor, $\tau$ is the conformal time; greek (latin) indeces
run from 0 to 3 (1 to 3); dots indicate differentiation in respect to
$\tau$. Around the energy scale $\Lambda_{QCD}$ (quark--hadron transition), 
we shall take \cite{6}
\begin{equation}
m(T,\phi) \simeq 0.1\, m_o(\phi) \left( \Lambda_{QCD} \over T 
\right)^{3.8} 
\label{eq:m2}
\end{equation}
with $m_o(\phi) = m_\pi f_\pi / \phi$. At $T <\sim 0.3$--$0.2\, 
 \Lambda_{QCD}$, $m(T,\phi)$ shall already approach its
low--$T$ behavior $m_o(\phi)$. 
The equations of motion 
then read
\begin{equation}
\ddot \theta + 2\left({\dot a \over a}+{\dot \phi \over \phi}\right) 
\dot \theta + m^2 a^2 \sin \theta = 0~,
\label{eq:m3}
\end{equation}
\begin{equation}
\ddot \phi + 2 {\dot a \over a}  \dot \phi + 
a^2  V'(\phi) = \phi\, \dot \theta^2 ,
\label{eq:m4}
\end{equation}
and will be mostly used with $\sin \theta \simeq \theta$. Then,
energy densities
$\rho_{\theta,\phi} = \rho_{\theta,\phi;kin} + \rho_{\theta, \phi; pot}$
 and pressures $p_{\theta,\phi} = \rho_{\theta,\phi;kin} -
\rho_{\theta, \phi;pot}$,  under the condition $\theta \ll 1$, are
obtainable from
\begin{eqnarray}	
\rho_{\theta,kin} = {\phi^2 \over 2 a^2} \dot \theta^2~,~~
\rho_{\theta,pot} =  {m^2(T,\phi) \over 2} \phi^2 \theta~,~~
\nonumber
\\
\rho_{\phi,kin} = {\dot \phi^2 \over 2 a^2} ~,~~
\rho_{\phi,pot} = V(\phi)~.~~~~~~~
\label{eq:kp}
\end{eqnarray}	

\section{The case of SUGRA potential}

When $\theta$ undergoes many (nearly) harmonic oscillations within a Hubble
time, $\langle \rho_{\theta,kin} \rangle \simeq 
\langle \rho_{\theta,pot} \rangle$ and $\langle p_\theta \rangle$ vanishes.
Under such condition, using
eqs.~(\ref{eq:m3}),(\ref{eq:m4}),(\ref{eq:kp}), 
it is easy to see that
\begin{equation}
\dot \rho_\theta + 3{\dot a \over a} \rho_\theta = {\dot m \over m}
 \rho_\theta
~,~~~~~~~~~~~ \dot \rho_\phi + 3 {\dot a \over a} (\rho_\phi+p_\phi)
= - {\dot m \over m}  \rho_\theta ~.
\label{eq:m7}
\end{equation}
When $m$ is given by Eq.~(\ref{eq:m2}), $\dot m/m
= -\dot \phi/\phi - 3.8\, \dot T/T$. 
At $T \simeq 0$, instead,  $\dot m / m \simeq -\dot \phi/\phi$.
Here below, the indices $\theta$, $\phi$ will be replaced by DM, DE.
Eqs.~(\ref{eq:m7}) clearly show
an exchange of energy between DM and DE.
Let us notice that the former Eq.~(\ref{eq:m7}) can be formally 
integrated, yielding $\rho_{DM} \propto m/a^3$. In particular, 
this law holds at $T \ll \Lambda_{QCD}$, and then 
\begin{equation}
\rho_{DM} a^3 \phi \simeq {\rm const.},
\label{eq:m8}
\end{equation}
so that the usual behavior $\rho_{DM} \propto a^{-3} $ is modified
by the energy outflow from DM to DE.

Let us now assume that the potential reads
\begin{equation}
V(\phi) = {\Lambda^{\alpha+4} \over \phi^\alpha} \exp (4 \pi\phi^2/m_p^2)
\label{eq:l1}
\end{equation}
and does not depend on $\theta$; in the radiation
dominated era, it admits the tracker solution
\begin{equation}
\phi^{\alpha+2} = g_\alpha \Lambda^{\alpha+4} a^2 \tau^2 ~,
\label{eq:l2}
\end{equation}
with $g_\alpha = \alpha (\alpha+2)^2/4(\alpha+6)$. 
This solution holds until we approach the quark--hadron
transition. Then, in Eq.~(\ref{eq:m4}), the term $\phi \dot \theta^2$ 
due to the DE--DM coupling, exceeds $a^2 V'$ and we
enter a different tracking regime. This is shown in detail in Fig.~1,
\begin{figure}[t]
\includegraphics[height=9.5truecm,angle=0.]{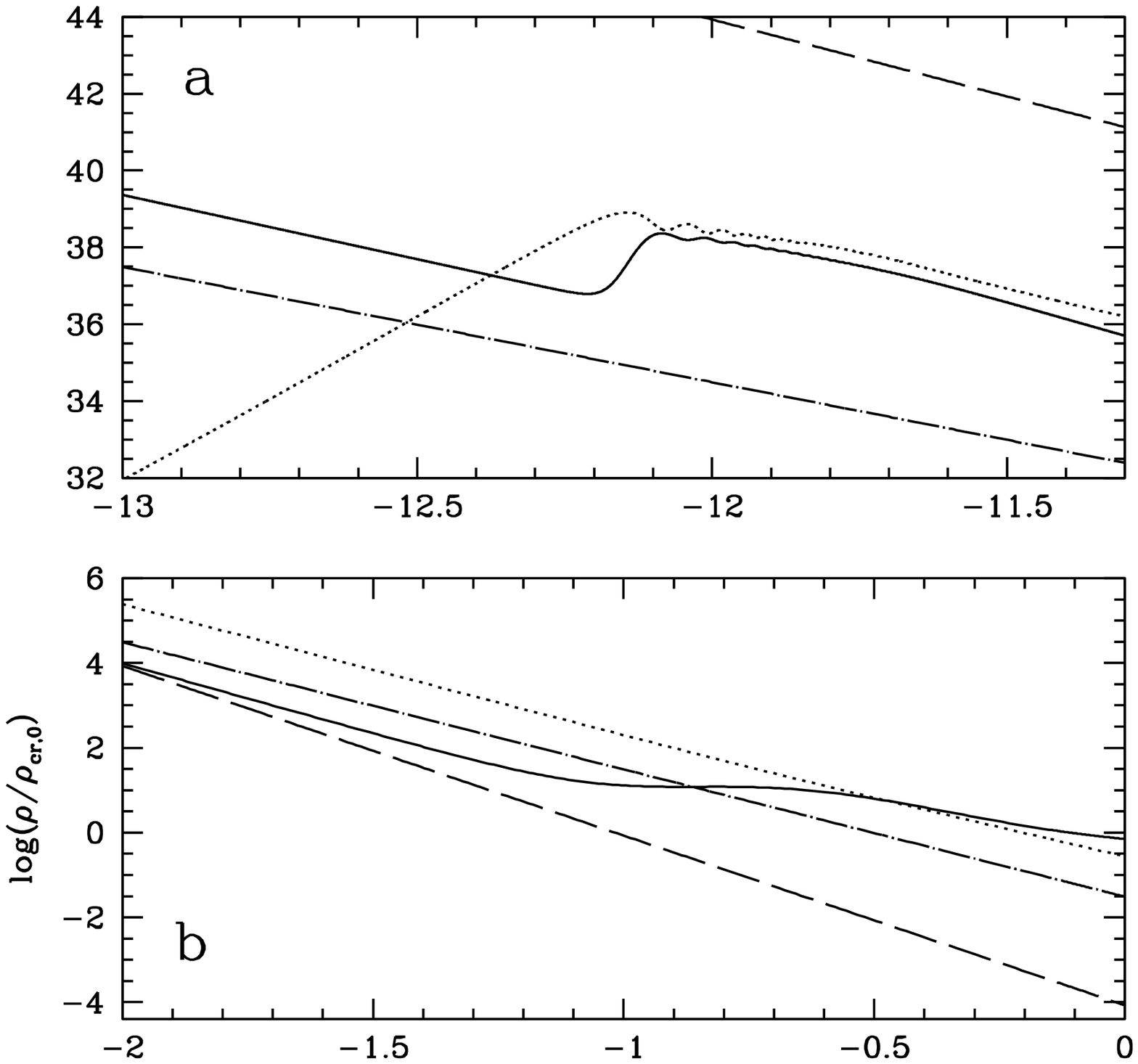}
\vskip-1.4truecm
\includegraphics[height=9.5truecm,angle=0.]{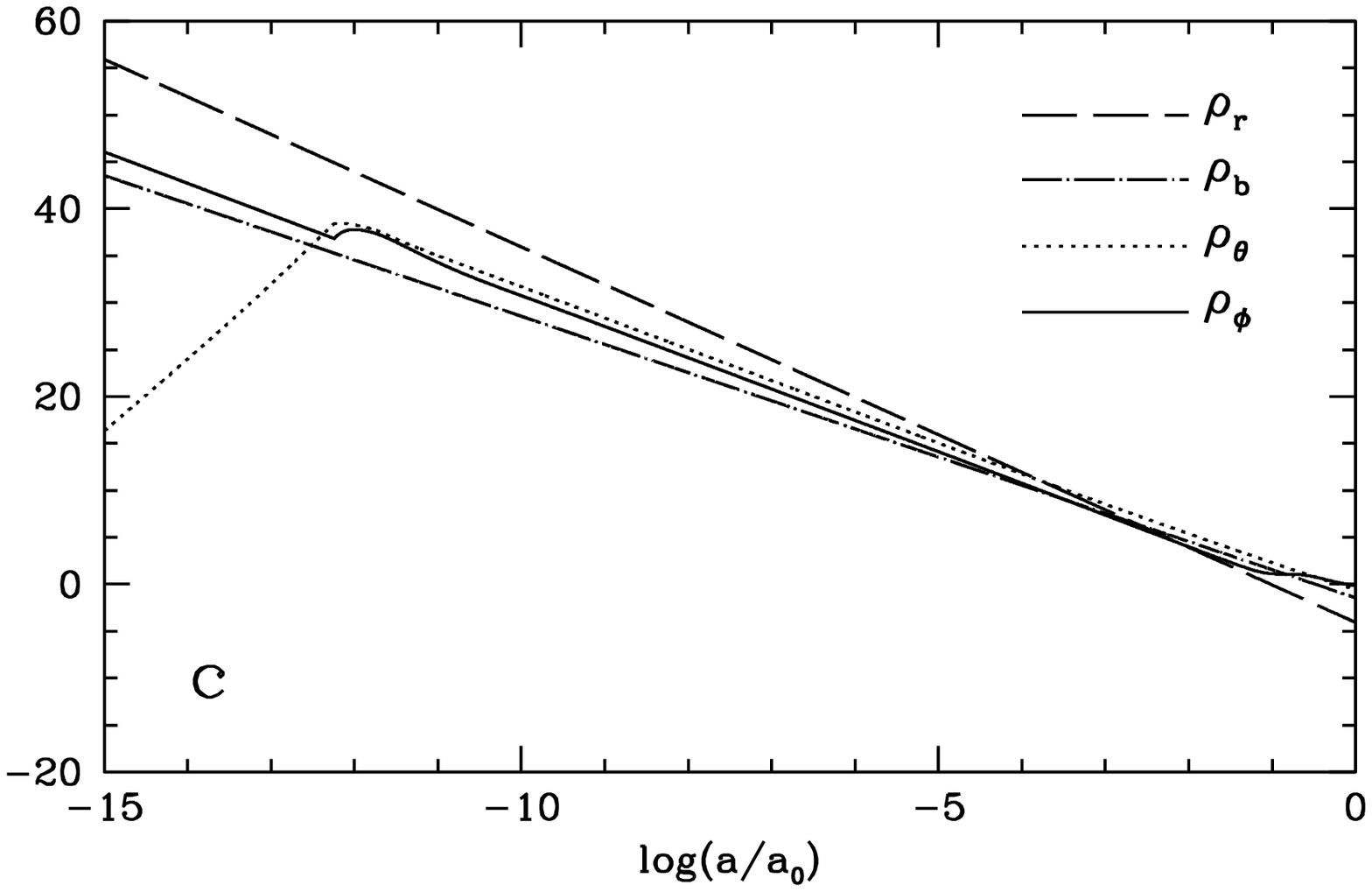}
\vskip-3.5truecm
\caption{Densities of the different components vs.~the
scale factor $a$. Fig.~1(a) magnifies the onset of the oscillation regime.
Fig.~1(b) shows the low--$z$ behavior. Fig.~1(c) is a landscape
picture of the whole evolution. All abscissas are $\log(a/a_0)$.}
\label{fig:rho}
\end{figure}
obtained for matter (baryon)
density parameters $\Omega_{m} = 0.3$ ($\Omega_b = 0.03$) and
$h = 0.7$ (Hubble constant in units of 100 km/s/Mpc). In particular, 
Fig.~1(a) shows the transition between these tracking regimes.
Fig.~1(b) then shows the low--$z$ behavior ($1+z=1/a$), since DE 
density exceeds radiation and then gradually overcomes baryons 
(at $z \sim 10$) and DM (at $z \simeq 3$). 
Fig.~1(c) is a landscape behavior of all components, down to $a=1$.
Notice, in particular, the $a$ dependence of $\rho_{DM}$, 
occurring according to Eq.~(\ref{eq:m8}). In Fig.~2 
\begin{figure}[t]
\includegraphics[height=9.5truecm,angle=0.]{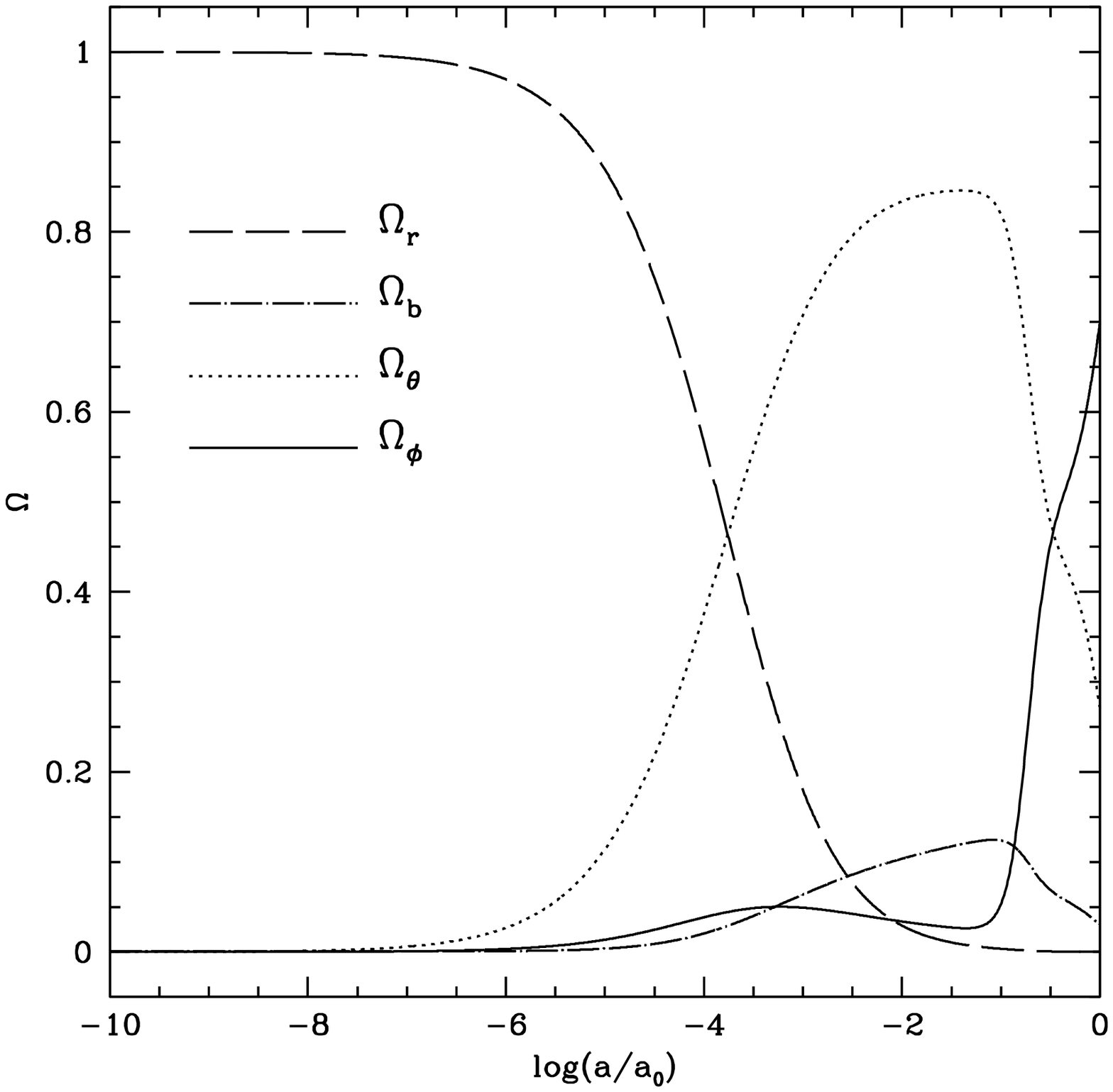}
\vskip-0.8truecm
\caption{Density parameters $\Omega_{r,~b,~\theta,~\phi}$
(radiation, baryons, DM, DE) vs.~the scale factor $a$.}
\label{fig:omega}
\end{figure}
we show the related behaviors of the density parameters 
$\Omega_i$ ($i = r,~b,~\theta,~\phi$, i.e.
radiation, baryons, DM, DE).

In general, once the density parameter $\Omega_{DE}$ (at $z=0$) 
is assigned, a model with dynamical (coupled or uncoupled) DE
is not yet univocally determined. For instance, the potential
(\ref{eq:l1}) depends on  the parameters $\alpha$ and $\Lambda$ 
and one of them can still be arbitrarily fixed. Other potentials 
show similar features.

In the present case such arbitrariness no longer exists.
Let us follow the behavior of $\rho_{DM}$, backwards in time,
until the approximation $\theta \ll 1$ no longer applies. 
This moment must approximately coincide
with the time when $\theta$ enters the oscillation regime.
This occurs when
\begin{equation}
2(\dot a/a + \dot \phi/\phi) \simeq m(T,\phi)\, a
\label{eq:k1}
\end{equation}
(see Eq.~\ref{eq:m3}). At that time, according to Eq.~(\ref{eq:m8}),
which is marginally valid up to there, and taking $\theta = 1$,
\begin{equation}
\rho_{DM} \simeq \rho_{o,DM} {\phi_o \over \phi(a)} {1 \over a^3}
\simeq m^2[T(a),\phi(a)]\, \phi^2(a) ~.
\label{eq:k2}
\end{equation}
The system made by eqs.$\, $(\ref{eq:k1}),(\ref{eq:k2}), owing to 
Eq.~(\ref{eq:l2}), yields the scale factor $a_h$ when fluctuations 
start and the value of $\Lambda$ in the potential (\ref{eq:l1}),
as soon as $\rho_{o,DM}$ (the present density of DM) is assigned.
\vglue 0.1truecm

The plots shown in the previous section, drawn for $\Omega_{DM}
= 0.27$, are obtained for $\Lambda \simeq 1.5 \cdot 10^{10}$GeV, 
as is required by eqs.$\, $(\ref{eq:k1}),(\ref{eq:k2}). In this
case $a_h \sim 10^{-13}$.
When $\Omega_{DM}$ goes from 0.2 to 0.4, $\log_{10}(\Lambda/
{\rm GeV})$ (almost) linearly runs from 10.05 to 10.39$\, $
and $a_h$ steadily lays at the eve of the quark--hadron transition.
A model with DE and DM given by a single complex field,
based on SUGRA potential, therefore bears a precise prediction on the
scale $\Lambda$, for the observational $\Omega_{DM}$ range.
In turn, we can say that, if the soft breaking of super--symmetries
occurred at a scale slightly above $10^{10}$GeV, $\Omega_{DE} \sim
0.3$ is a natural consequence.

\section{Evolution of inhomogeneities}

Besides of predicting fair ratios between the world components,
a viable model should also allow the formation of structures
in the world. This matter will be treated in detail in a forthcoming
paper. Let us however outline that the model treated here belongs to 
the class of coupled DE models
treated by Amendola \cite{7}, with a time--dependent coupling.
In fact, for small $\theta$'s, the r.h.s. of eqs.~(\ref{eq:m4}),
after averaging over cosmological times, reads $ C(\phi) \langle 
\rho_\theta \rangle a^2$
with $C(\phi ) = 1/\phi$. Similarly, in eqs.~(\ref{eq:m7}),
which are already averaged, the r.h.s. are  $\pm C(\phi)
\, \dot\phi\, \rho_\theta$ 
($C$ is the DE--DM coupling introduced in \cite{7}). Let us also outline 
that Fig.~2 shows a $\phi$--$MDE$ phase, typical of this class 
of models, after matter--radiation equivalence,
as the kinetic energy of DE is non--neglegible
during the matter--dominated era.

By solving the fluctuation equations in \cite{7}, with the
above $C(\phi)$, we find the behavior shown in Fig.~3~(top).
\begin{figure}[t]
\includegraphics[height=9.5truecm,angle=0.]{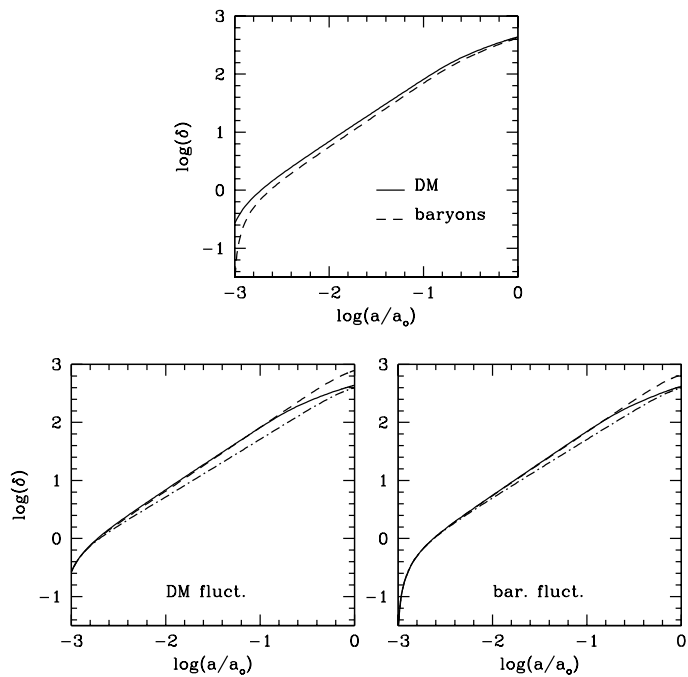}
\vskip-0.8truecm
\caption{Time evolution of DM and baryon fluctuations.
The top figure shows DM and baryon fluctuation evolution in this model.
The two bottom figures compare DM and baryon fluctuation evolutions in
this model (solid curve), $\Lambda$CDM (dot-dashed),
coupled DE (dashed).}
\label{fig:w}
\end{figure}
Figs.~3~(bottom) compare
fluctuation evolutions in this model (solid curves), with those in an
analogous $\Lambda$CDM model (dot--dashed curves) and in a coupled DE model
with constant coupling $C=0.25\, \sqrt{8\pi G} \simeq\langle C(\phi) \rangle$ 
(dashed curves). These plots show that the
linear growth factor, from recombination to now,
is significantly smaller than in coupled DE models
with constant coupling and, more significantly, is quite close to 
$\Lambda$CDM. The essential differences from $\Lambda$CDM are that:
(i) objects should form earlier; (ii) baryon fluctuations keep
below DM fluctuations until very recently.

\section{Discussion}

The first evidences of DM date 70 years ago, but
its non--baryonic nature became compulsory in the Seventies, 
when BBNS and CMBR anisotropies were studied. DE is younger,
but is now required both by SNIa data \cite{8}, as well as by CMBR
and deep galaxy data \cite{9,10}. Axions have been candidate
DM since the late Seventies, although various studies,
as well as the occurrence of the SN 1987a, have finally
constrained the PQ scale around values $F_{PQ} \sim 10^{12}$GeV.
Contributions to DM from topological singularities
(cosmic string and walls) have also narrowed the constraints to
$F_{PQ}$ \cite{11}. Here they
were disregarded and could cause shifts in
quantitative predictions. We shall deepen this point in further work.

The fact that scalar fields can yield both DM or DE,
just changing an exponent in the potential,
stimulated the work of various authors. A potential
like (\ref{eq:l1}) was considered in 
{\it spintessence} models \cite{12}. According to the choice of
parameters, $\Phi$ yields either DM or DE.

On the contrary, in this note we deal with the possibility that
$\Phi$ accounts for {\it both} DE and DM, and that the strong--$CP$ 
problem is simultaneously solved. As in the PQ model, the angle $\theta$
in Eq.~(\ref{eq:n1}) is turned into a dynamical variable, i.e.
into the phase of a complex scalar field $\Phi$, and is gradually
driven to approach zero, by our cosmic epoch. 
Residual $\theta$ oscillations, yielding axions, account for DM.
The critical time for the onset of coherent axion oscillations
occurs at the eve of the quark--hadron transition,
because of the rapid increase of $m(T,\phi)$. Here $\phi$
replaces the constant $F_{PQ}$ scale.
This stage is illustrated by Figs.~4(a), 4(b)
\begin{figure}[t]
\includegraphics[height=9.5truecm,angle=0.]{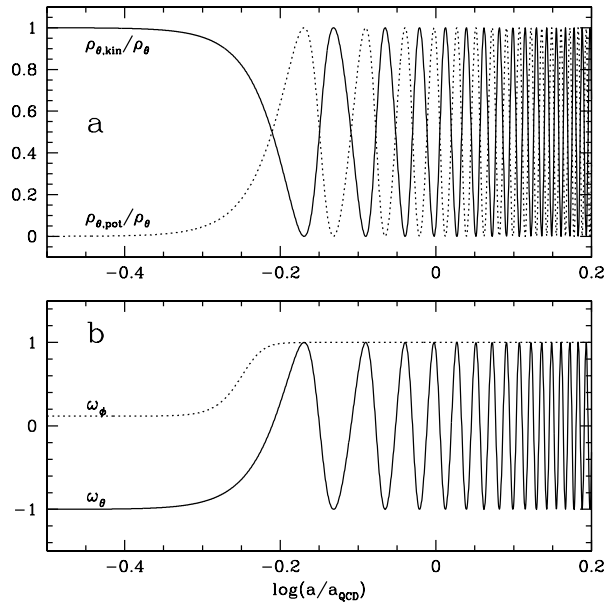}
\vskip-0.8truecm
\caption{The onset of coherent axion oscillations,
at the eve of the quark--hadron transition,
due to the increase of $m(T,\phi)$, causes the behaviors of 
$\rho_{\theta,pot}$, $\rho_{\theta,kin}$ (a) and
$\omega_{\phi,\theta} = p_{\phi,\theta}/\rho_{\phi,\theta}$ (b)
shown here.}
\label{fig:w}
\end{figure}
where the behaviors of $\rho_{\theta,pot}$, $\rho_{\theta,kin}$, and
$\omega_{\phi,\theta} = p_{\phi,\theta}/\rho_{\phi,\theta}$, 
are plotted. 

The novel features of this model arise because
the expectation value of $\phi$ is not
a constant $F_{PQ}$, but evolves over cosmological times; in a sense,
we let $F_{PQ}$ evolve so to yield DE.
Such evolution modifies the friction term in
Eq.~(\ref{eq:m3}). The damping of $\theta$ oscillations is
therefore greater and $\theta$ oscillations are smaller today.
Further, accordingly to Eq.~(\ref{eq:m2}) the axion mass (or
the oscillation frequency), which varies fairly rapidly
during the formation of $\bar q q$ condensate, continues to
evolve, over cosmological scales, due to the evolution
of the $\phi$ field.
The (low--)$z$ dependence of $m_o(\phi)$ is shown in Fig.~5.
\begin{figure}[t]
\includegraphics[height=9.5truecm,angle=0.]{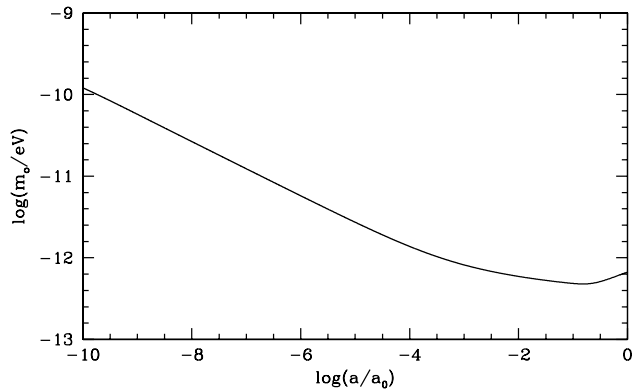}
\vskip-3.8truecm
\caption{$\phi$ variations cause a dependence of the effective axion
mass on scale factor $a$, which is shown here.}
\label{fig:mass}
\end{figure}
We draw the reader's attention on the rebounce at $z \sim 10$,
whose implications on halo formation could be critical \cite{13}.

Constraints on PQ axions came from $z=0$ observations, which
must be fulfilled by the same $F_{PQ}$ scale, fulfilling also cosmological
requirement. Here, $\phi$ attains
values $\sim m_p$ today, so that most these constraints should be naturally
satisfied. This matter, as well as the question of a direct axion
detection, will be deepened in further work.

Let us outline that the choice of a SUGRA potential is arbitrary
and could be replaced by other potentials, perhaps better approaching
data at $z=0$. However, using this potential allows to appreciate
the conceptual echonomy in this approach. In the PQ approach,
the $F_{PQ}$ scale is assumed. Here, once
$\Lambda$ is set in a physically significant range, around the
quark--hadron transition, $\phi$ naturally passes through 
values enabling to solve the strong--$CP$ problem and later
naturally grows to values providing fair amounts of DM and DE.

\begin{acknowledgments}
Acknowledgments. Luciano Girardello, Antonio Masiero and Federico Piazza are
gratefully thanked for useful discussions.	
\end{acknowledgments}


{}
\end{document}